\documentclass[twocolumn,pra]{revtex4-1}
\usepackage[T1]{fontenc}
\usepackage[utf8]{inputenc}
\usepackage{amsmath,amssymb}
\usepackage{color}
\usepackage{graphicx}
\usepackage{amssymb}
\begin{document}
\title{Coherent tunneling and negative differential conductivity in graphene-hBN-graphene   heterostructure.}
\author{ Luis  Brey}
\affiliation{Instituto de Ciencia de Materiales de Madrid, CSIC, 28049 Cantoblanco, Spain}
\email{brey@icmm.csic.es}

\date{\today}

\begin{abstract}
We address the tunneling current in a graphene-hBN-graphene heterostructure as function of the twisting between the crystals.
The twisting induces a modulation of the hopping amplitude  between the graphene layers,  that provides the extra momentum necessary to satisfy  momentum and energy conservation and to activate
coherent tunneling between the graphene electrodes.  
Conservation rules  limit the tunneling to  states with wavevectors  lying  at the conic curves defined by the intersection of two Dirac cones shifted in momentum and energy.  There is a critical voltage where the intersection is a straight line, and the joint density of states presents a maximum. This reflects in a peak in the tunneling current and in a negative differential conductivity.

 \end{abstract}
\maketitle

{\it Introduction.} The same techniques used for obtaining graphene layers\cite{Novo_2004} can also be  applied to obtain two-dimensional (2D) crystal structures of highly anisotropic materials  as hexagonal Boron Nitride (hBN)\cite{Dean_2010} or transition metal dichalcogenides\cite{Mak_2010}.
Once isolated, 
atomic layers of different 2D crystals can  be reassembled layer by layer to create heterostructures with the designed electrical  properties\cite{Geim_2013}. 
In this direction, 
recently   graphene-hBN-graphene\cite{Lee_2011,Britnell_2012,Britnell_2013}
and graphene-WS$_2$\cite{Georgiou_2013} heterostructures have been realized and proved as prototype graphene based field-effect tunneling transistors.  At high voltages the graphene-hBN-graphene structure shows a negative differencial conductance\cite{Britnell_2013} that has potential applications for logic devices.


Conservation of energy and momentum prevents  finite voltage coherent tunneling between 2D-electron gases with
circular symmetric dispersion. Coherent tunneling only  occurs when  the Fermi surfaces  of the electron gases are closely aligned\cite{Eisenstein_1991}.

In this work we show that in graphene-hBN-graphene (G-BN-G) heterostructures,  
the lattice mismatch  between graphene and hBN  induces an unavoidable twisting  and a spatial modulation of the hopping amplitude between the graphene electrodes.  This  translates  into  a coherent tunneling current between the graphene layers and a negative differential conductivity.
We find that even in the case of perfect crystal arrangement  between the graphene layers, the always present misalignment between graphene and hBN, makes possible coherent tunneling between the graphene electrodes.

{\it Geometry and Model.} We consider a trilayer structure consisting of  top (T) and  botton (B) graphene monolayers separated by L monolayers of 
hBN. 
T and B graphene layers are rotated angles $\theta _T$ and $\theta _B$ respectively with respect the central hBN layers, and they have a lattice parameter mismatch $\delta$=1.8$\%$ with  hBN.
For small twisting angles, the tunneling amplitude between the  layers varies over distances much larger than the  lattice constant and electronic states in Dirac points  $\bf K$ and $\bf K'$ are effectively decoupled. Therefore  we  describe each valley separately.
Near the Dirac point, ${\bf K} = (k_D ,0)$ with $k_D$=$ \frac {4 \pi}{3a}$, the Hamiltonians  for the T and B graphene layers are\cite{Guinea_2009},
\begin{equation}
h^{T(B)}_{\rm k} = \hbar v_F  \! \left( \begin{array}{cc}
0 &k e ^{i ( \theta _{\bf k} -  \theta _{T(B)})}  \\
ke ^{-i ( \theta _{\bf k} - \theta _{T(B)})}  & 0 \\
\end{array} \right) 
\label{Dirac}
\end{equation}
here $v_F$ is the graphene Dirac velocity, ${\bf k}$ is the momentum measured from the layer's Dirac point and $\theta _{\bf k}$ is the angle formed by the momentum with the $x$-axis. 
Hamiltonian $h^{T(B)}_{\rm k}$ acts on the amplitude of the wavefunction on 
the sublattices, A and B, of the  graphene layer T(B). 
The electronic structure of each hBN monolayer is described by a  gapful Dirac-like Hamiltonian that acts on the B and N atomic basis, 
\begin{equation}
h^{BN}_{\rm k} =  \! \left( \begin{array}{cc}
\Delta _1 & \hbar v_{BN} ke ^{i  \theta _{\bf k}  }  \\
\hbar v_{BN}ke ^{-i  \theta _{\bf k} }  & -\Delta _2 \\
\end{array} \right) \end{equation}
where $v_{BN}$ describes the in-plane hopping amplitude  between B and N atoms, $\Delta _1$+$\Delta_2$ is the energy gap of hBN and
$\Delta_1$ is the band-offset  of the conduction  band, Boron-like,  of hBN with respect the graphene Dirac point.   
 The different hBN layers are vertically ordered in an eclipse way  and the atoms  are coupled by a vertical  hopping $\gamma _{BN}$.  This vertical order is a consequence of the bond polarity in hBN.

Top and bottom  graphene  layers are coupled with the first and last hBN layers  through the spatially modulated hopping matrices
$ V(\theta _T, \delta)$ and $   V(\theta _B, \delta)$ respectively, that in the low twisting angle limit have the form\cite{Lopes_2007,Bistritzer2011,Kinderman_2012}
\begin{equation}
V(\theta,\delta) = \frac {\hat t } 3 \sum _{i=1,3} T _i  \, e ^{-i \bf {q}_i (\theta,\delta) \bf{r}} ,
\label{tmod}
\end{equation}
with
$
T_1$=$ \left( \begin{array}{cc}
1&1 \\
1&1
\end{array} \right )$,  
$T_2$=$\left( \begin{array}{cc}
\eta ^* &1 \\
\eta & \eta ^* \end{array} \right ) $,
$T_3$=$\left( \begin{array}{cc}
\eta &1 \\
\eta ^* &\eta \end{array} \right )$, $\hat t$=$\left ( \begin{array} {cc} t_{CB} & 0 \\ 0 &  t_{CN} \end{array} \right )$ and  $\eta$=$ e^ {i \frac {2 \pi} 3}$. Being   $t_{CB}$ and $t_{CN}$  the C to B and C to N hopping amplitudes  respectively.
The  hopping matrices $T _i$  do not depend on  geometrical factors. All the information on $\delta$ and $\theta$  is in the ${\bf q_i}$'s. 
\begin{eqnarray}
\bf{q}_1 (\theta,\delta) & = & k_D \left (\delta,-\theta \right ),  \nonumber \\
\bf{q}_2 (\theta,\delta) & = & 
k_D \left (- \frac {\sqrt{3}} 2 \theta + \frac 1 2 \delta,- \frac 1 2 \theta- \frac {\sqrt{3}} 2 \delta \right ),\nonumber  \\
\bf{q}_3 (\theta,\delta) & = & 
k_D \left ( \frac  {\sqrt{3}} 2 \theta + \frac 1 2 \delta,- \frac 1 2\theta  + \frac {\sqrt{3}} 2 \delta \right ) \, \, 
\label{wvq}
\end{eqnarray}
The three wavevectors ${\bf q}_i$ have the same modulus 
and define a periodic hexagonal modulation of the hopping amplitude. 
This periodicity describes the spatial  distribution of the stacking of the graphene C atoms with the B and N atoms of hBN.

{\it Effective Hamiltonian.}
We obtain an effective bilayer graphene  Hamiltonian by integrating out the orbital degree of freedom in the hBN layer,
\begin{equation}
\hat H_{\bf k} = \left( \begin{array}{cc}
h ^T  & 0  \\
0 & h ^B  \\
\end{array}\right )+\left( \begin{array}{cc}
0& \hat V  \\
\hat V ^{\dagger} &0 \\
\end{array}\right )
\end{equation}
where,
\begin{eqnarray}
\hat V  & =&\hat t \,  V(\theta _T,\delta ) \left ( H _{\bf k} ^{BN} \right ) ^{-1} V(-\theta_B,-\delta) \, \hat t \, ,
\end{eqnarray}
and $H_{\bf k} ^{BN}$ is the Hamiltonian of the L layers hBN slab.
For wavevectors, ${\bf k}$, of the order of the separation between the Dirac points of the T and B graphene layers, $|{\bf q}_i |$, the  diagonal terms $\Delta_1$ and $\Delta_2$ are the leading  contributions in the hBN Hamiltonian $h^{BN}_{\bf k}$.  For those momenta it is a very good approximation to set $v_{BN}$=0 in $ h _{\bf k} ^{BN} $, resulting in the following  T to B graphene tunneling modulation,
\begin{equation}
\hat V = \frac 1 9  \sum _{i,j=1,3} \hat  T _{i,j} \,  e ^{ i {\bf G} _{i,j}(\theta_T,\theta_B)  {\bf r}}
\end{equation}
with 
\begin{equation}
 {\bf G} _{i,j}(\theta_T,\theta_B)={\bf q} _i (\theta _T,\delta)+{\bf q}_j (-\theta _B,-\delta) \, 
\end{equation}
and
\begin{equation}
{\cal T}_{i,j} = \frac {\gamma _ {BN} ^{L-1}}{(\Delta _1 \Delta _2 ) ^L } \, \hat t \, \,  T _i \left ( \begin{array}{cc} \Delta _2 ^L & 0 \\ 0 & \Delta _1 ^L \end{array} \right )  T _j \, \, \hat t
\end{equation}
The three  tunneling processes linking T graphene with  hBN,  combine with the  three connecting hBN with  B graphene. This results in
nine Fourier components of the tunneling modulation between T and B graphene layers. 
The three {\it diagonal} wavevectors $\{{\bf G}_{ii}\}$ have a modulus $G_d$=$k_D|\theta _T - \theta _B|$ and  vanish when T and B layers are aligned. The six {\it non-diagonal} transfer momenta have modulus $G_{nd}$=$k_D \sqrt{\theta_T ^2 +\theta _B ^2 +\theta _T \theta_B}$. 
Therefore, even when the two graphene layers are perfectly aligned, the misalignment with the central hBN layer makes possible tunneling processes between
the graphene electrodes. Note that because the T and B graphene layers have the same lattice parameter,
the wavevectors $\{{\bf G}_{ij}\}$ are independent on the graphene-hBN lattice mismatch, $\delta$.

{\it Tunneling Current  in G-BN-G heterostructure.}
In presence of an applied voltage $V$, between the T and B graphene electrodes, the tunneling current   can be obtained in linear response theory with the tunneling term treated as the perturbation\cite{Mahan_book}.
\begin{widetext}
\begin{equation}
I(V) = \frac e {\hbar} g_s g_v 
\sum_{\substack{{\bf k} ,\{i,j\} \\ \alpha,\beta}}
 |t_{\alpha,\beta} ({\bf k},{\bf k}+{\bf G}_{ij})|^2 \int _{-\infty} ^{+\infty} \frac {d \omega} {2 \pi}
A _{\alpha} ({\bf k}, \hbar \omega )A _{\beta} ({\bf k}+ {\bf G}_{ij}, \hbar \omega +eV) 
\left [  n_F (\hbar \omega)-n_F (\hbar \omega +eV)   \right ]\, ,
\label{IV}
\end{equation}
\end{widetext}
where $g_s$=$2$ and $g_v$=$2$ account for the spin and valley degeneracy respectively, $\alpha =\pm$ is the band index,
$n_F(\epsilon)$=$\left [ \exp((\epsilon -E_F)/k_B T)+1\right ] ^{-1}$ is the Fermi factor, $A _{\alpha} ({\bf k}, \hbar \omega )$ is the graphene spectral function for band $\alpha$,  and $t_{\alpha,\beta} ({\bf k},{\bf k}+{\bf G}_{ij})$ is the tunneling matrix element between states in the T and B unperturbed graphene layers,
\begin{equation}
t_{\alpha,\beta} ({\bf k},{\bf k}+{\bf G}_{ij})=\phi_{\alpha} ^* ({\bf k}) \, {\cal T} _{i,j}  \, \phi_{\beta} ({\bf k}+{\bf G}_{ij}) \, ,
\label{tme}
\end{equation}
being
$
\phi _{\alpha} ({\bf k})$=$ \frac 1 {\sqrt 2}
\left ( \begin{array}{c} 1 \\ \alpha e ^{i \theta _{\bf k}}
\end{array} \right )
$ the Dirac hamiltonian eigenfunction  with momentum ${\bf k}$ and energy $\alpha \hbar v_F k$.
In the previous expressions $E_F$ is the Fermi energy of the T and B graphene layers that we consider equally doped.
In the one electron picture, the spectral function should be proportional to a delta function, in our calculations $A_{\alpha}$ is  approximated by a Lorentzian function centered  on the band energy, $\alpha \hbar v_F k$, and with an half  width at half maximum  $ \hbar /\tau$.

The tunneling processes corresponding to different  transfer 
wavevectors ${\bf G}_{i,j}$ contribute independently to the current, and  because the circular symmetry of the  graphene band structure, their contribution to the current only depends on their modulus  $|{\bf G}_{i,j}|$. Therefore the relevant quantities or the tunneling current are the two modulus $G_d$ and $G_{nd}$.

It is important to note that there is   current between the two graphene layers because they are rotated with respect the central hBN layer. 
In systems with circular symmetric band structure,  only the  presence of the unavoidable disorder or phonons make possible  the observation of  finite voltage incoherent tunneling between two 2D electron gases separated by a barrier.
On the contrary, in the trilayer G-BN-G heterostructure, the spatial
modulation of the hopping amplitude between T and B layers, provides an extra wavevector that make possible
the conservation of momentum and energy in the coherent tunneling process.

 \begin{figure}[htbp]
\includegraphics[width=\columnwidth]{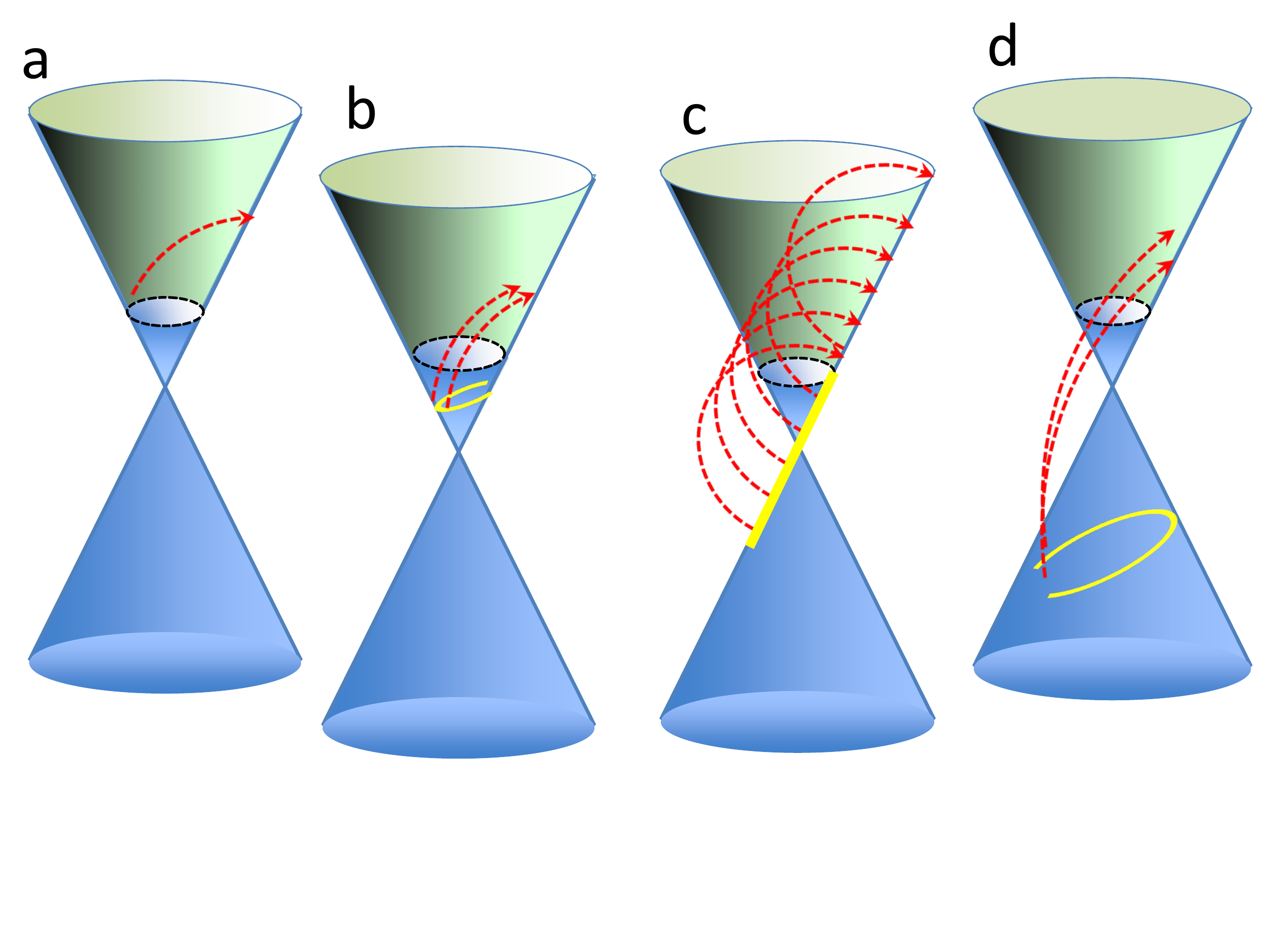}
\caption{(Color online)Schematic representation of  tunneling processes occurring in the G-hBN-G heterostructure.
Blue and green regions mark occupied and empty states respectively.
The arrows  indicate a tunneling event  from the T layer  (initial point) to the B layer (end point).  These points are shifted in energy by $eV$.
Energy and momentum  conservation laws define permitted  initial  curves in ${\bf k}$ space which are plotted  in yellow.
a) At the minimum voltage for tunneling, only a point in ${\bf k}$-space can tunnel,  b) by increasing the voltage the conservation curve is an hyperbola that resides in the
conduction band, c) at a critical voltage the hiperbola collapses in a straight line and a peak in the  joint density of states  occurs. d) At larger voltages, 
the straight line becomes an ellipse, now residing  in the valence band.
}
\label{Diagram}
\end{figure}

It is possible to get some insight on the different tunneling contributions analyzing   the momentum and energy conservation, together with the Fermi occupation of the T and B layers.   In the linear 
regime, the conduction is different from zero only if the relation $\hbar v_F |{\bf G} _{i,j}|< E_F$ is satisfied.  That implies finite conductance for twisting angles inside the regions  defined by the relations 
$\hbar v_F k_D|\theta_T +\theta _B|<E_F$ or $\hbar v_F k_D \sqrt{ \theta _T ^2 + \theta _B ^2 - \theta _T \theta _ B} < E_F$. 
In general, except for very small twist angles, it is appropriated  to  assume that both $G_{nd}$ and $G_d$ are smaller than $k_F$ and therefore in the linear regime the current is zero.

At finite voltage,
energy and momentum  conservation define a curve  in reciprocal space for the initial tunneling states in the top layer.
In general these curves are the conic sections defined by the intersection of two Dirac cones shifted a momentum ${\bf G}_{i,j}$ and an energy
$eV$. 
At small voltages,  
the tunneling  connects conduction  band states,
and the
permitted tunneling wavevectors  define an hyperbola. 
At larger voltages, electrons in the valence band of the top layer can tunnel to the conduction band states of the bottom layer and  the allowed momentums form an ellipse. Both hyperbola and ellipse lengths increase with the voltage, and  the current should increases
continuously with voltage. 
However, there is a
critical voltage where the hyperbola transforms to an ellipse adopting the form of a straight segment. 
At this critical voltage $V^{c}$=$\hbar v_F |{\bf G}_{i,j}|$ the cones intersect along two parallel lines  and there is a spike  in the joint density of states  that translates in a  peak in the tunneling current. This peak is the origin of the  negative differential conductivity in this heterostructure.

The states defined by these conic curves are further limited by the Fermi occupation. That imposes a  minimum voltage $V^{min}$=$\hbar v_F(|{\bf G}_{i,j}|-2k_F)$ for the existence of   tunneling current. The Fermi occupation also  constrains  the wavevectors  of the states that tunnel at $V^c$ to be in the interval  
 $ |{\bf G}_{i,j}|$-$k_F<k<k_F$.


\begin{figure}[htbp]
\includegraphics[width=\columnwidth]{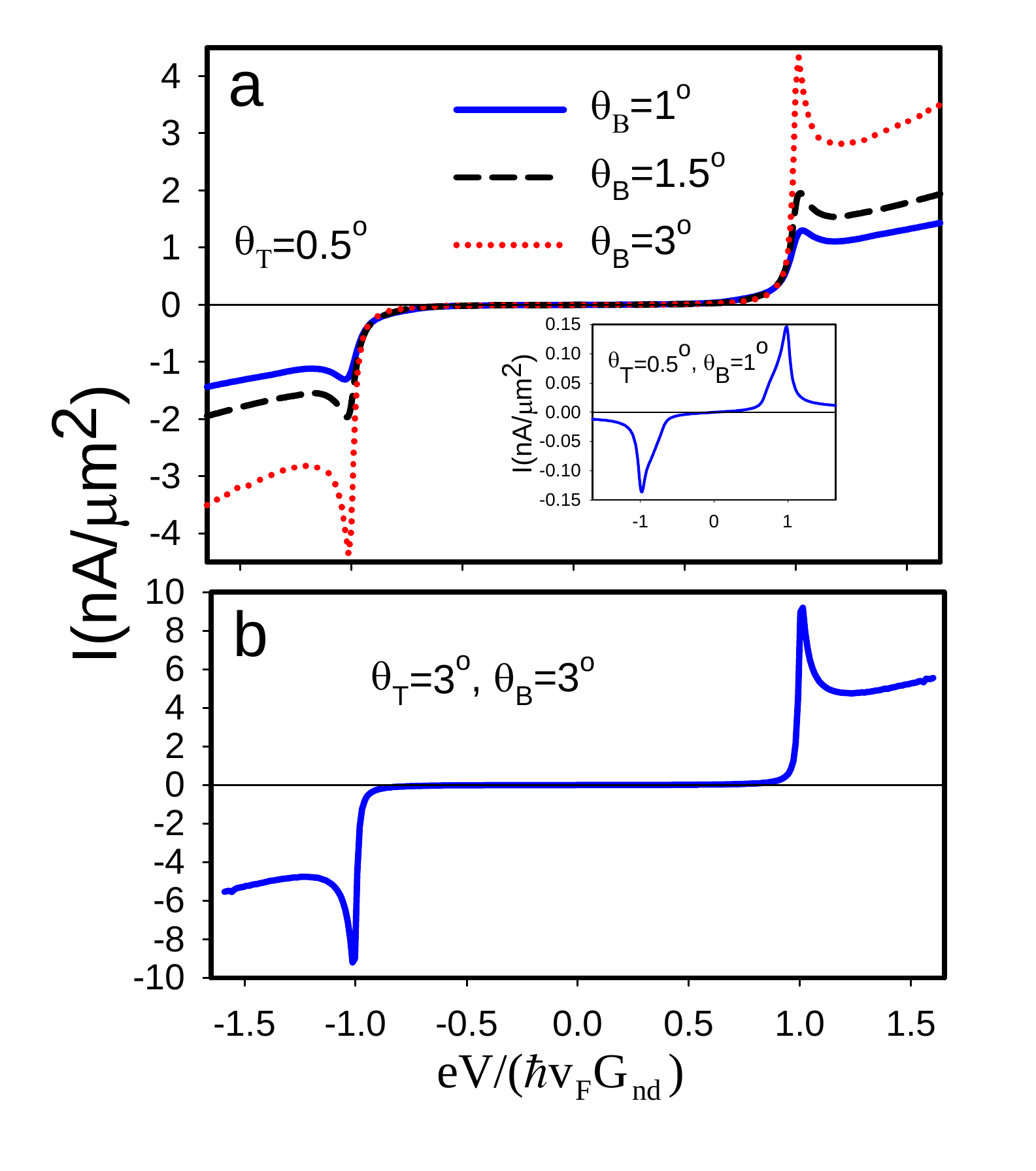}
\caption{(Color online)Non-linear current for a graphene-hBN-graphene heterostructure.  The density of electrons in both layers is
$n$=$5\times 10 ^{12}$cm$^{-2}$. In panel a) we fix the top layer twist angle to 
$\theta_T$=0.5$^o$ and plot the current for different rotation angles of the bottom layer
$\theta _B$.  Both angles are measured with respect the central hBN layer.  
In the inset we show the conduction to conduction  ($\alpha$=$+$ to $\beta$=$+$) contribution to the current. 
In panel b) both graphene layers are rotated the same angle
$\theta_T$=$\theta_B$=3$^o$. The peak in the $I(V)$ indicates that coherent tunneling and negative differential conductivity can occur even
when  both graphene layers are fully aligned, provided there is a twisting with the hBN layer.
In the calculation we use the band structure parameters\cite{Jung_2013} $\Delta _1$=3.33eV, $\Delta _2$=1.49eV, $t_{CB}$=0.432eV and $t_{CN}$=0.29eV
 and a  value $\hbar / \tau $=2.5meV.}
\label{Figure2}
\end{figure}

{\it Numerical Results.}
A precise description of the tunneling current requires the evaluation of the tunneling matrix elements, Eq.\ref{tme}, which depends on the numerical values of the tight-binding parameters.
We have used the band structure parameters recently obtained from {\it ab initio} calculations by Jung {\it et al} \cite{Jung_2013}.
In Fig.\ref{Figure2} we plot the current for a G-hBN-G heterostructure with different twisting angles.  We obtain that at small angles,  the tunneling processes associated with the transfer of  diagonal momentums ${\bf G}_ {i,i}$  have a practically null contribution to the current. The main tunneling current is associated with the non diagonal momentum and therefore
we measure
the bias voltages in units of  $\hbar v_F G _{nd}$.

In the inset of the upper part of Fig.\ref{Figure2} we show the intraband contribution to the current.  The interband contribution is activated at 
voltage $\hbar v_F(|{\bf G}_{nd}|-2k_F)$ and is zero for voltages larger than $V^c$=$\hbar v_F|{\bf G}_{nd}|)$.  
For $V>V^c$ all the tunneling current has its origin in  interband processes.  Both 
inter and intraband tunneling show a strong peak at this critical voltage. As discussed above this peak is related to a big increase of the joint density of states occurring at this voltage.

At $V^c$  the interband peak is much stronger than the intraband one. 
This is because in the  intraband tunneling  only states with wavevectors in a segment of length $k_F$ contribute the current.
However  for interband tunneling the number of wavevector contributing to tunneling is  proportional to $G_{nd}-k_F$.
Then, the strong peak in the $I(V)$ curve is due to valence band to conduction band tunneling processes. 
In panel a) of Fig.\ref{Figure2}, we see that as the twisting angles become larger, the value of the momentum transfer increases and with it the intensity of the negative differential peak.
Finally the numerical results confirm that  the negative differential conductivity peak exists even when both graphene layers are fully aligned, lower panel of Fig.\ref{Figure2}.

{\it In-plane magnetic field.}
A magnetic field applied parallel to the graphene layers 
affects differently to the distinct  Fourier components of the interlayer tunneling. Then we expect that the magnetic field splits the negative differential conductivity peak. The experimental observation of this effect  would be a definitive  indication of the coherent nature of the tunneling. 

The magnetic field ${\bf B}_{\parallel}$=$B_{\parallel} (\cos \beta, \sin \beta,0)$ is described in the Landau gauge, ${\bf A}$=$B_{\parallel} (\sin \beta \, z, -\cos \beta \, z,0)$. For isolated graphene layers an in-plane magnetic field shifts the position of the Brillouin zones, and its effect can be cancelled by distinct gauge transformation for the two graphene sheets.
Thus, in absence of tunneling the magnetic field has not physical relevance. When electrons can hope between the graphene layers, the motion of the carriers perpendicular to the magnetic field is affected by ${\bf B}_{\parallel}$
\cite{Hayden_1991,Brey_1988,Falko_1991}.
 and the  shift in the $k$-space reflects in a shift in the tunneling wavevectors,
\begin{equation}
{\bf G}_{i,j}  \, {\Large{\rightarrow}}  \, {\bf G}_{i,j}- \frac d {\ell _{\parallel} ^2} (\sin \beta, -\cos \beta )
\end{equation}
being $\ell _{\parallel} = \sqrt {\frac {\hbar c}{e B_{\parallel}}}$ the magnetic length and $d$ the separation between the graphene ayers.
The modulus of the new wavevectors ${\bf G}_{i,j}$ depends both on the magnitude of $B_{\parallel}$ and on the its in-plane orientation. 
 
The position of  the peak in the $I(V)$ curve is  determined by the modulus of the transfer wavevector. For $B_{\parallel}$=0  the six  non diagonal 
wavevectors have the same modulus, $G_{nd}$,  and only a peak  appears, see Fig.\ref{Figure2}. The magnetic field  modifies the modulus of the transfer wavevectors  and the peak in the $I(V)$ curve broadens and splits in presence of $B_{\parallel}$.

\begin{figure}[htbp]
\includegraphics[width=\columnwidth]{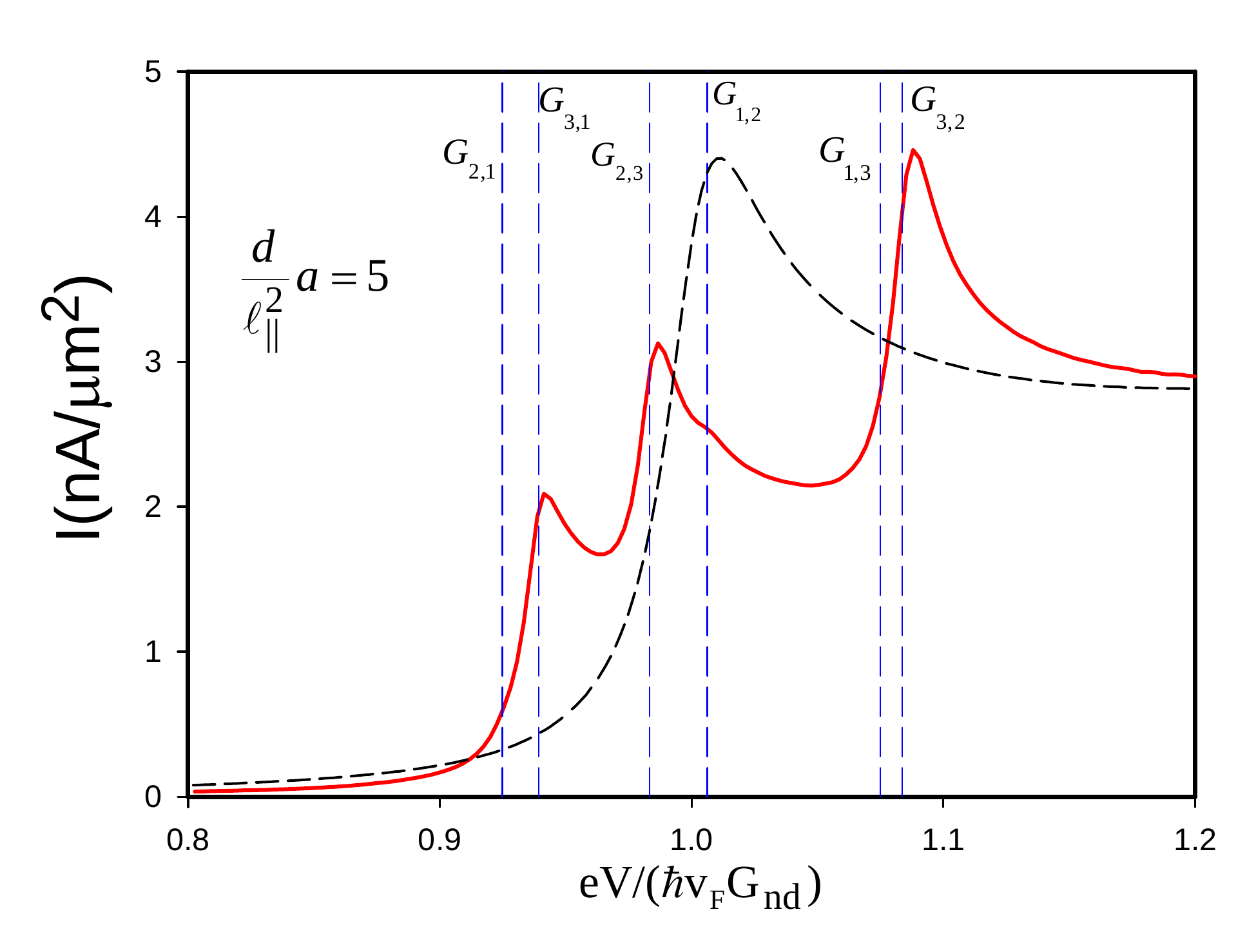}
\caption{(Color online)$I(V)$ curve or a G-BN-G structure with twisting angles $\theta _T$=-0.5$^o$ and $\theta _B$=3$^o$, 
in presence of  an in-plane magnetic field corresponding to
$  d {\ell _{\parallel} ^{-2}}a$=5, being $a$ the graphene lattice parameter.  The density of carriers in both layers is $n$=5$\times$10$^12$cm$^{-2}$. The current for $B_{\parallel}$=0 is  shown as a   black dashed line. Vertical lines indicate the critical voltages corresponding to  the 
different magnetic field modified
tunneling transfer wavevectors
${\bf G}_{i \ne j}$.  }
\label{B20}
\end{figure}

In Fig.\ref{B20} we plot the effect of $B_{\parallel}$ on the $I(V)$ peak, for a particular G-hBN-G heterostructure.
The negative differential peak splits in three clear peaks, corresponding to three different  transfer wavevectors. The other three wavevectors only produce small shoulders  only visible in derivates of the curve. The intensity and resolution  of the peaks depends on the tunneling amplitude, on the strength of $B_{\parallel} $, on the  modulus of the transfer momentum and on the in-plane orientation of the magnetic field.

In order to observe the effect of the magnetic field, the quantity $d \ell ^{-2}$ should be comparable to the value of the 
modulus of the momentum transfer $G_{nd}$. This can be achieved by  increasing the number of hBN layers or using a 
very strong magnetic field.  The results presented in Fig.\ref{B20} correspond to just one hBN layer, and the magnetic field corresponding to  $  d {\ell _{\parallel} ^{-2}}a$=5 is of the order of 80$T$.  By increasing the number of hBN layers the separation between graphene layer become larger  and the magnetic field required for observing the splitting of the peak should be more accessible.  


We note that recent field effect tunneling\cite{Britnell_2012} and negative differential conductance\cite{Britnell_2013} experiments in Gr-hBN-Gr heterostructures have been explained by assuming disorder induced momentum conservation  relaxation and therefore non-coherent tunneling.  Also, recently Feenstra {\it et al.}\cite{Feenstra_2012}
considered the tunneling between $n$- and $p$-doped   graphene layers separated by a dielectric barrier.
That work applied the transfer Hamiltonian formalism  to model   the tunneling  between missoriented graphene layers, 
and the information on the dielectric crystal structure is neglected.

In summary, we have studied the tunneling current between two graphene layers separated by a hBN layer. The twisting of  the layers induces a spatial modulation of the hopping amplitude between the graphene electrodes that provide  extra wavevectors to the tunneling process.
These  extra momenta make possible the conservation of  energy and momentum and  activates coherent tunneling. 
Because of the   Dirac-like linear dispersion of graphene, the wavevectors that conserve 
energy and momentum in the  tunneling process,  can be defined as the intersection  of two Dirac cones shifted in momentum and energy. 
At a critical voltage, the intersection conic curves  collapse 
in a straight segment, and there is a strong peak in the joint density of states and in the tunneling current.

When finishing this work we learnt about the experimental work of K.Novoselov {\it et al.} where possible signatures of negative differential conductance and coherent tunneling in G-hBN-G heterestructures were reported\cite{Novoselov_2014}.



\begin{acknowledgments}
This work has been partially supported by MEC-Spain under grant
FIS2012-33521. Luis Brey thanks C.Tejedor for a critical reading of the manuscript.
\end{acknowledgments}

%


\end{document}